\documentstyle[epsf,a4,12pt]{article}

\begin{document}

\begin{titlepage}

\vspace*{2.truecm}

\centerline{\Large \bf  Short-time Dynamic Behaviour of  }
\vskip 0.6truecm
\centerline{\Large \bf 
                        Critical XY Systems }
\vskip 0.6truecm

\vskip 2.0truecm
\centerline{\bf H.J. Luo, M. Schulz$^*$, L. Sch\"ulke, 
S. Trimper$^*$ and B. Zheng$^*$}
\vskip 0.2truecm

\vskip 0.2truecm
\centerline{Universit\"at -- GH Siegen, D -- 57068 Siegen, 
Germany}

\centerline{$^*$Universit\"at -- Halle, D -- 06099 Halle, Germany}

\vskip 2.5truecm

\abstract{ Using Monte Carlo methods, the short-time dynamic 
scaling behaviour of two-dimensional
critical  XY systems is investigated. 
Our results for the XY model show that there exists
universal scaling behaviour already in the short-time regime,
but the values of the dynamic exponent $z$ differ
for different initial conditions.
For the fully frustrated XY model, power law scaling behaviour is
also observed in the short-time regime.
 However, a violation of the standard scaling relation
between the exponents
is detected.
}

\vspace{0.3cm}


\vspace{0.1cm}


\end{titlepage}

Recently much progress has been made in critical dynamics.
It was discovered that universal scaling behaviour may emerge 
already
in the {\it macroscopic} short-time regime
\cite {jan89,hus89,sta92,li94,sch95,zhe98}.
Extensive Monte Carlo simulations show that 
the short-time dynamic scaling is not only conceptually
interesting but also practically important,
e.g. it leads to new ways for the
determination of the critical exponents and the 
critical temperature \cite 
{zhe98,luo98,li96,sch96,gra95,blu92,sta92}.

For critical systems with second order phase transitions,
comprehensive understanding has been achieved.
For a relaxational dynamic process of model A starting from
an ordered state, 
the scaling form is given, e.g.
for the $k$-th moment of the magnetization 
at the critical temperature, by \cite {zhe98,sta92}
\begin{equation}
M^{(k)}(t,L)=b^{-k(d-2+\eta)/2}M^{(k)}(b^{-z}t,b^{-1}L),
\label{e20}
\end{equation}
where $t$ is the time variable,  $L$ is the lattice size, 
$\eta$ and $z$ represent the standard static and dynamic
critical exponents. This scaling form looks similar
to that in the long-time regime but it is now assumed to hold
also in the macroscopic short-time regime after a microscopic 
time scale
$t_{mic}$.

For a relaxation process starting from a disordered state
with small or zero initial magnetization,
the scaling form for the $k$-th moment at the critical temperature
is found to be
\begin{equation}
M^{(k)}(t,L,m_{0})=b^{-k(d-2+\eta)/2}
M^{(k)}(b^{-z}t,b^{-1} L,
b^{x_{0}}m_{0}).
\label{e40}
\end{equation} 
Here it is important that a new independent 
critical exponent $x_0$ has been introduced 
to describe the dependence of the scaling behaviour on the 
initial magnetization. The exponent $x_0$ is 
the scaling dimension of the global magnetization
$M(t)$ and also of the magnetization density.
Therefore, even if $m_{0}=0$, $x_0$ still enters
observables related to the initial conditions.
For example, for sufficiently large
lattice size the auto-correlation 
 has a power law behaviour 
\begin{equation}
A(t) \equiv \frac{1}{L^d}\, \langle\sum_i s_i(0) s_i(t)\rangle 
\sim t^{-\lambda}
\label{e60}
\end{equation}
with $s_i$ being a spin variable.
The exponent $\lambda$ is related to $x_0$ by \cite {jan92}
\begin{equation}
\lambda=\frac{d}{z}-\theta\;, \qquad  
\theta=\left(x_0 -\frac{d-2+\eta}{2} \right)\frac{1}{z}\;.
\label{e80}
\end{equation}
In general, for arbitrary initial magnetization
$m_0$ between $0$ and $1$ a generalized scaling form
can be written down.

Renormalization group calculations for the $O(N)$ vector
model show that the static exponents $\eta$ 
and the dynamic exponent $z$ in the scaling forms
 (\ref {e40}) and (\ref {e60}) take the same values
 as in equilibrium or in the
 long-time regime of the dynamic evolution where they are 
defined.
 This is also confirmed by Monte Carlo simulations for various
 critical magnetic systems with second order
 phase transitions. Furthermore,
 numerical results indicate in good accuracy
 that the values of the exponents $\eta$ and $z$
 are independent of the initial conditions,
 i.e. the values of the exponents in scaling form (\ref {e20})
 are the same as in scaling forms (\ref {e40}) and (\ref {e60}).
 This prominent property has been used to extract all the
 dynamic and static 
 exponents from the short-time dynamic scaling forms.
 Since the measurements are carried out in the short-time 
regime,
 the short-time dynamic approach is free of critical slowing down.
 One may believe that the scaling forms 
(\ref {e20}), (\ref {e40}) and (\ref {e60}) hold also
in the crossover regime where the time $t$ is not small
but also not asymptotically large.

In this letter we investigate numerically whether
critical systems with a Koster\-litz-Thouless
phase transition show also a clean scenario 
as stated above. For this purpose,
we should answer the following questions: 
at the first, whether there exists
universal scaling behaviour in the short-time regime;
secondly, whether the dynamic exponent $z$ and the
static exponent $\eta$ in the short-time scaling forms
(\ref {e20}), (\ref {e40}) and (\ref {e60}) are
independent of the initial conditions and take the same values
as in equilibrium or in
the long-time regime where they are defined\footnote 
{In principle, even if there exist the short-time
scaling form as in Eqs. (\ref {e20}), (\ref {e40}) and (\ref {e60}),
the dynamic exponent $z$ and even
the static exponents could be initial condition dependent and
different from those in equilibrium
since they are defined in different fixed points
of the time $t$ and different initial conditions.};
 finally whether the scaling relations between
the exponents for the second order phase transitions,
e.g. that between $x_0$ and $\lambda$ in Eq.~(\ref {e80}),
 hold also for
Kosterlitz-Thouless phase transitions. 

As examples, let us consider the two-dimensional
classical XY model and the fully frustrated XY (FFXY) model.
The dynamic process starting from an ordered state
has been investigated with Monte Carlo methods
at the Kosterlitz-Thouless
phase transition $T_{KT}$ and below \cite {luo97a}.
Nice power law scaling behaviour was observed,
and the static exponent $\eta$ extracted from scaling form
(\ref {e20}) is consistent with that measured in equilibrium.
The dynamic exponent $z$ is a constant for all temperatures
at and below $T_{KT}$ and slightly smaller than $2.0$,
but within errors it coincides with the theoretical prediction
$z=2$ \cite {rut95}.

For the dynamic process starting
from a disordered state, however,
the situation is complicated. For a quench to {\it zero}
temperature, it was suggested some years ago
that the dynamic scaling is violated \cite {blu94}.
However, recent numerical simulations show that
at least for a quench to not so low temperatures
the magnetization undergoes a universal power law
initial increase if the initial magnetization
is small but non-zero \cite {oka97,luo98a}.
Therefore, a comprehensive understanding of
the short-time behaviour of this dynamic process
is important and necessary. 

Our strategy is to simulate a dynamic relaxational
process starting from a disordered state
with zero magnetization and to
measure the auto-correlation and the second moment of the 
magnetization.
Then from the scaling forms (\ref {e40}) and (\ref {e60}),
we obtain the static exponent $\eta$ and the dynamic exponent
$z$ and compare them with those measured from
the scaling form (\ref {e20}). 
Indeed, to simulate a dynamic process starting from a disordered 
state with zero magnetization
is more difficult than that
from an ordered state.
This is very probably 
due to the effect of the vortices. We need very large lattices
and update the system to fairly long time $t$. 

The XY model and the FFXY model in two dimensions
can be defined by the Hamiltonian
\begin{equation}
H=K  \sum_{<ij>} f_{ij}\  \vec S_i \cdot \vec S_j\ ,
\label{e100}
\end{equation}
where $\vec S_i = (S_{i,x},S_{i,y})$ is a planar unit vector at site
$i$, and
the sum extends over the nearest neighbours.
In our notation, $K$ is just  the inverse temperature.
Here $f_{ij}$ take
the values $+1$ or $-1$, depending on the model.
For the XY model, $f_{ij} = 1$ on all links.
A simple realization of the FFXY model is by taking
$f_{ij}=-1$ on half of the vertical links (negative links)
and  $+1$ on the others (positive links) \cite {sch89}.

In this paper, only the dynamics of model A is concerned.
The dynamics of model A is a relaxational dynamics
without energy and magnetization conservation.
Starting from a completely random configuration,
we update the system at the temperature $T_{KT}$
or below with the standard Metropolis algorithm. 
The trial state of each spin is randomly taken 
in the unit circle since the acceptance rate is high
in the short-time regime of the dynamic evolution.
Updating is stopped at Monte Carlo time step
$t=1500$. The procedure is repeated with another
initial configuration and different random numbers.
We have tested that for updating time $t=1500$,
a big lattice size like $L=512$ is needed.
An updating time $t=1500$ is necessary for reliable
measurements of the observables since the microscopic time
scale is here around $t_{mic} \sim 100$ or more.
For the average a total of $800$ samples has been taken. 
Errors are estimated
by dividing the samples into four groups.

We measure the auto-correlation
\begin{equation}
A(t) \equiv \frac{1}{L^d}\, \langle\sum_i \vec S_i(0) \cdot \vec S_i(t)
\rangle 
\label{e120}
\end{equation}
and the second moment of the magnetization
\begin{equation}
M^{(2)}(t) \equiv 
\frac{1}{L^{2d}}\, \left\langle\left[\sum_i \vec S_i(t) 
\right]^2\right\rangle\;. 
\label{e140}
\end{equation}
For the FFXY model, the second moment must be calculated
in four sublattices separately, and the results are summed up
after average.
Keeping in mind that spatial correlation
is very small in the short-time regime
of the dynamic evolution,
from the finite size scaling in Eq.~(\ref {e40})
one may easily deduce the short-time power law behaviour for
the second moment
\begin{equation}
M^{(2)}(t) \sim t^y 
\label{e160}
\end{equation}
with
\begin{equation}
y = (2-\eta)/z. 
\label{e180}
\end{equation}

\begin{figure}[h]
\epsfysize=6.5cm
\epsfclipoff
\fboxsep=0pt
\setlength{\unitlength}{0.6cm}
\begin{picture}(9,9)(0,0)
\put(0.,-0.5){{\epsffile{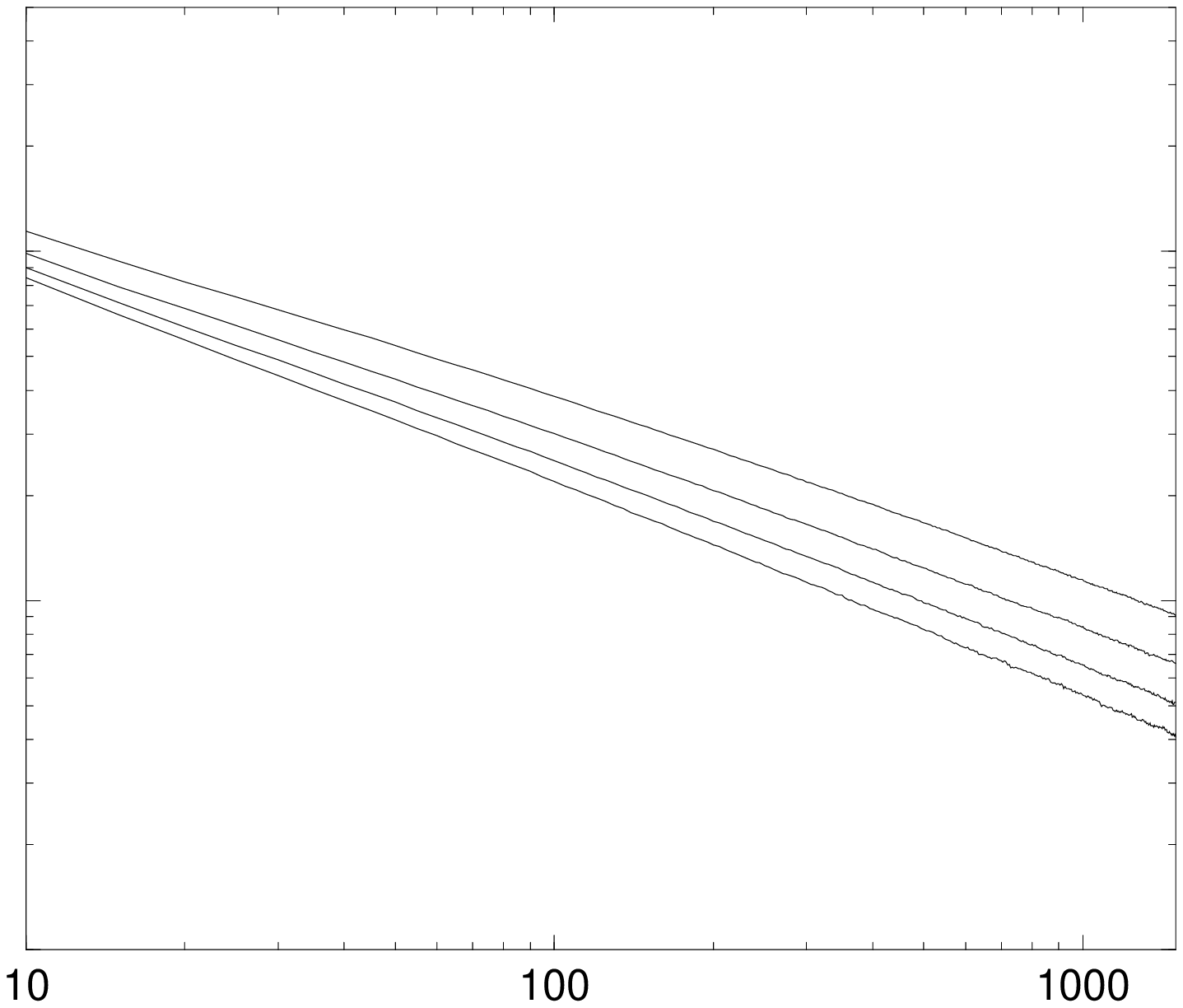 }}}
\put(1.0,8){\makebox(0,0){\footnotesize $A(t)$}}
\put(9.0,0.5){\makebox(0,0){\footnotesize $t$}}
\put(9.0,8.){\makebox(0,0){\footnotesize $(a)\ XY$}}
\epsfysize=6.5cm
\put(13.,-0.5){{\epsffile{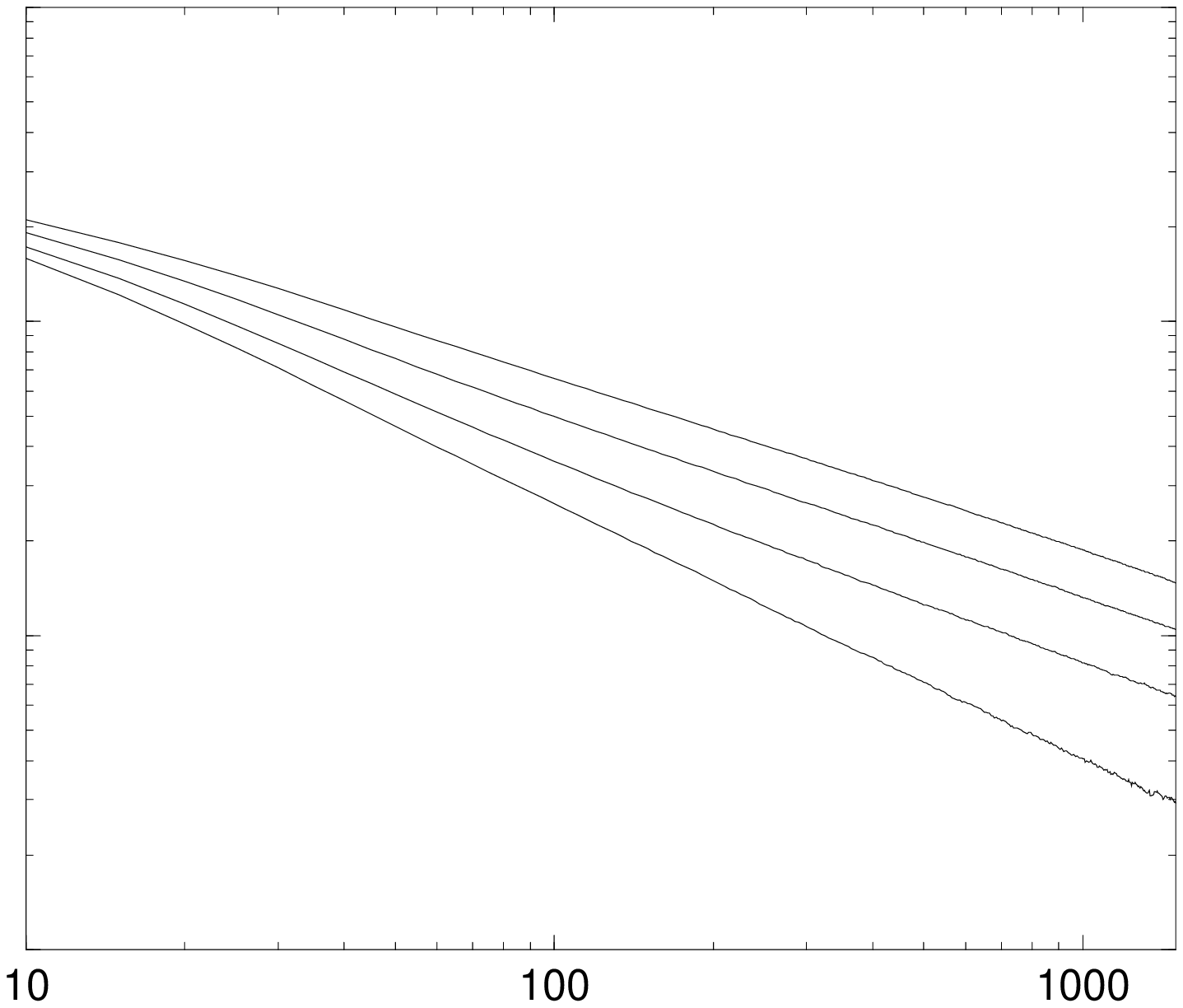}}}
\put(22.0,8.){\makebox(0,0){\footnotesize $(b)\ FFXY$}}
\end{picture}
\caption{Auto-correlation 
in log-log scale (a) for the XY model at the temperatures
$T=0.90$, $0.86$, $0.80$ and $0.70$ (from below);
 (b) for the FFXY model at the temperatures
$T=0.44$, $0.40$, $0.35$ and $0.30$ (from below).}
\label{f1}
\end{figure}

In Fig.~\ref {f1}, the auto-correlation
for both the XY and the FFXY model is displayed
in log-log scale for different temperatures at and below
$T_{KT}$. After a certain microscopic time scale $t_{mic}$,
power law behaviour is clearly observed.
Careful analysis shows that for temperature $T$ not so far from
$T_{KT}$, $t_{mic} \sim 100 - 200$ but as $T$ decreases,
$t_{mic}$ gradually increases.
Actually, this tendency can also be seen roughly by eyes
from the figure. For example,
for the XY model at $T=0.70$, $t_{mic} \sim 300 - 400$.
This increase of $t_{mic}$ for lower temperatures
was also noticed in the measurements of the
critical initial increase of the magnetization
\cite {oka97,luo98a}. This phenomenon is somehow 
understandable
since at low temperatures the configuration
tends to be frozen. As the temperature $T$ approaches
zero, we may believe that $t_{mic}$ becomes
considerable big. Numerical simulations
of the short-time behaviour for very low temperatures
are much more difficult. It requests
long updating time $t$ and correspondingly
large lattices. This might explain
the violation of dynamic scaling observed for
a quench to zero temperature \cite {blu94}.

From the slopes of the curves in Fig.~\ref {f1},
we measure the exponent $\lambda$.
The results are given in Tables~\ref{t1} and \ref{t2}.
As the temperature decreases, $\lambda$
becomes smaller. 
We should mention that the errors here are only those
estimated by dividing the total samples into four groups.
Other errors as the fluctuation in time direction
and a possible remaining effect of $t_{mic}$
have not been taken into account\footnote {The effect of 
$t_{mic}$ 
may sometimes be 
a kind of correction to the scaling which may not disappear
suddenly, e.g. when the correction obeys power law.}.
In equilibrium,
it is well known that critical slowing down
is much more severe in the XY and the FFXY model 
than in the Ising model. Numerical measurements
in the XY systems are very difficult.
However, the situation is completely different for
short-time dynamic measurements. 
For a lattice size $L=512$ and updating time $t=1500$,
we already obtain rather accurate results
with a total number of $800$ samples. 

\begin{figure}[h]
\epsfysize=6.5cm
\epsfclipoff
\fboxsep=0pt
\setlength{\unitlength}{0.6cm}
\begin{picture}(9,9)(0,0)
\put(0.,-0.5){{\epsffile{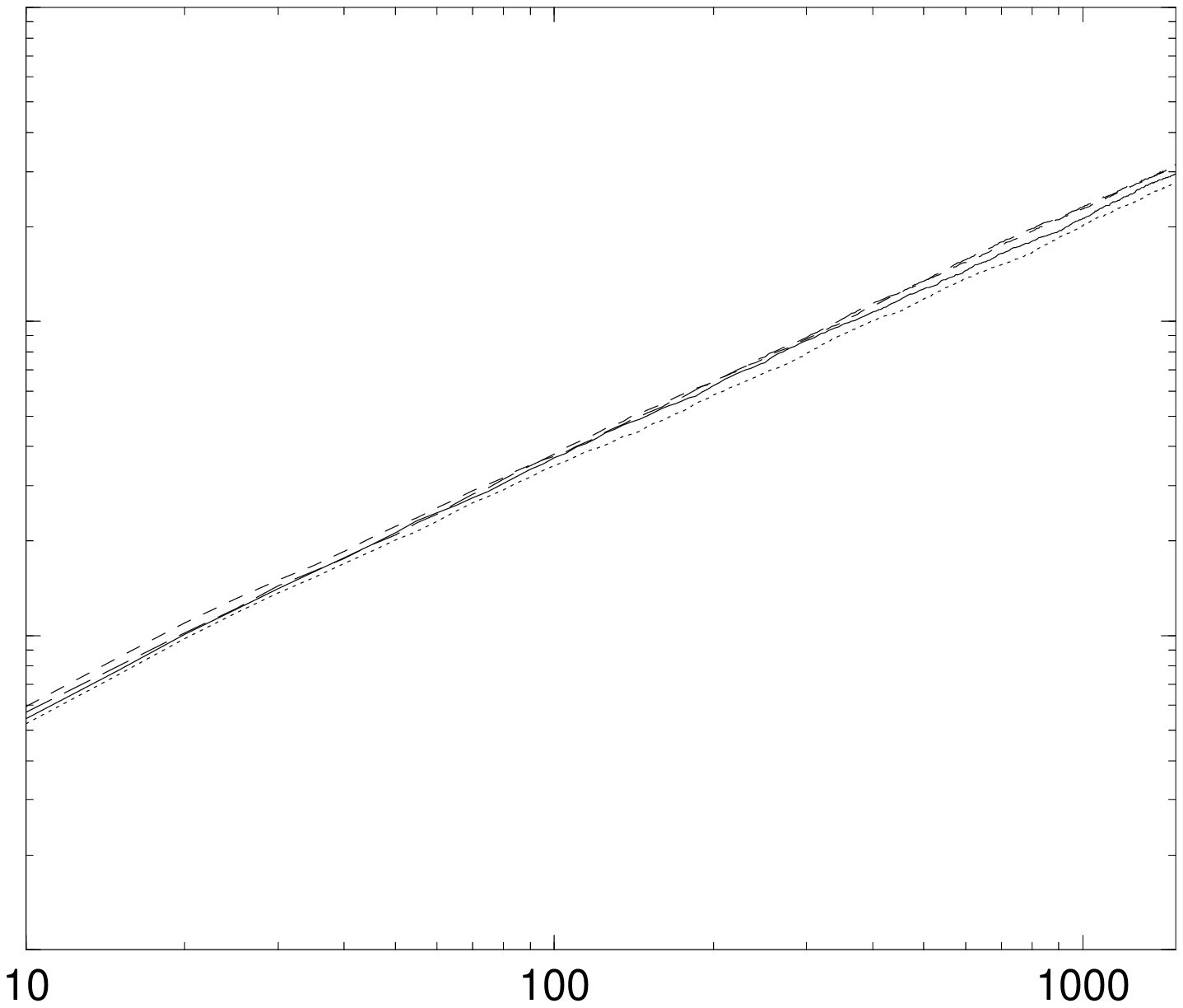 }}}
\put(1.0,8){\makebox(0,0){\footnotesize $M^{(2)}$}}
\put(9.0,0.5){\makebox(0,0){\footnotesize $t$}}
\put(5.,8.){\makebox(0,0){\footnotesize $(a)\ XY$}}
\epsfysize=6.5cm
\put(13.,-0.5){{\epsffile{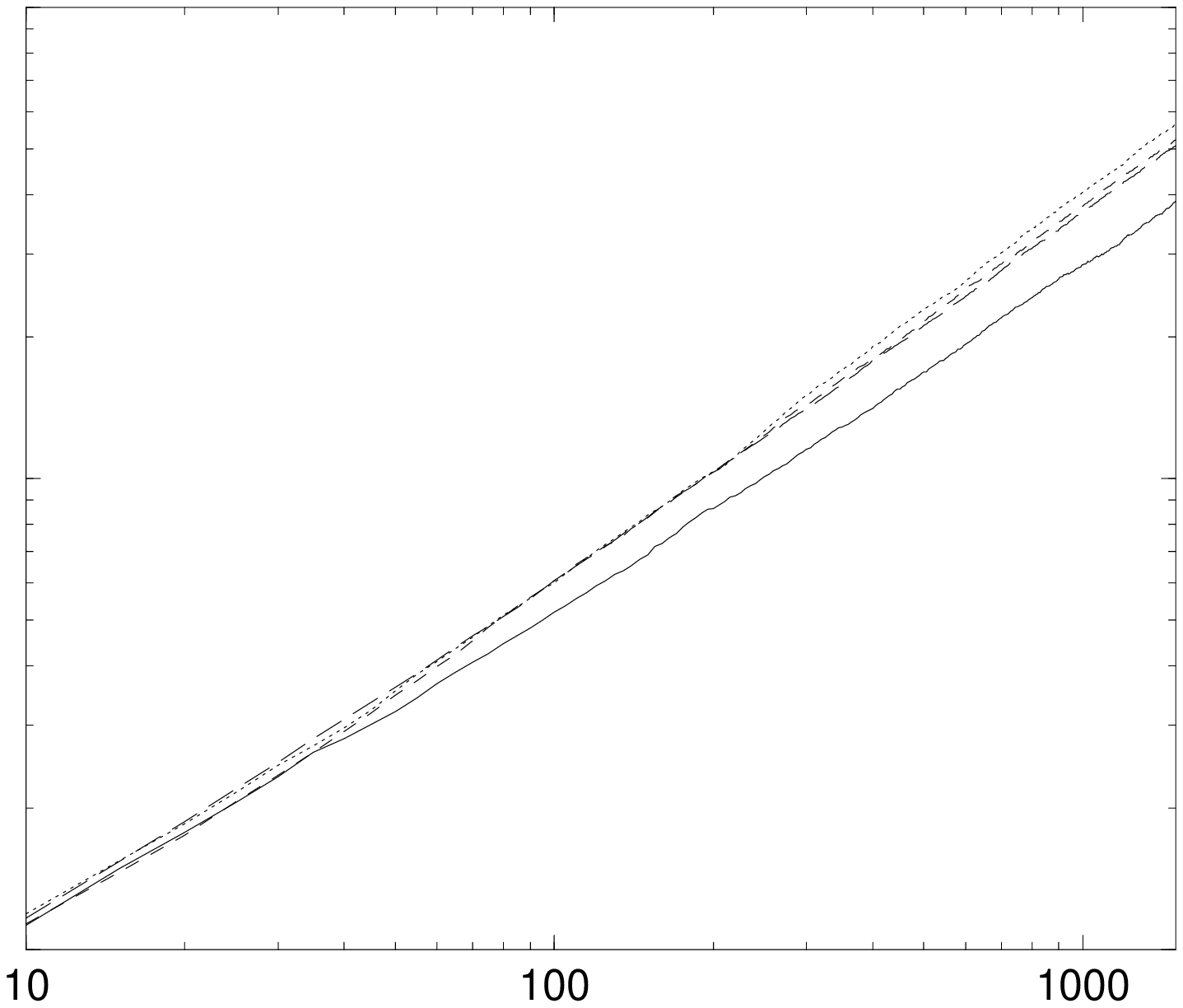}}}
\put(18.0,8.){\makebox(0,0){\footnotesize $(b)\ FFXY$}}
\end{picture}
\caption{Second moment 
in log-log scale (a) for the XY model with the solid, dotted,
dashed and long dashed lines for the temperatures
$T=0.90$, $0.86$, $0.80$ and $0.70$ respectively;
 (b) for the FFXY model with the solid, dotted,
dashed and long dashed lines for the temperatures
$T=0.44$, $0.40$, $0.35$ and $0.30$ respectively.}
\label{f2}
\end{figure}

In Fig.~\ref {f2}, the second moment
for both the XY and FFXY model is displayed
in log-log scale for different temperatures at and below
$T_{KT}$. Again we observe power law behaviour
even though the fluctuations here
are apparently larger than those of the auto-correlation.
From the slopes of the curves,
we measure the exponent $y$.
The results are also listed in Tables~\ref {t1} and \ref {t2}.
Interestingly, for the FFXY model the exponent $y$ obviously
first increases and then decreases as the temperature goes 
down.
This might be related to the rapid drop of the exponent 
$\lambda$
near $T_{KT}$.

In Tables~\ref {t1} and \ref{t2},
the static exponent $\eta$ is estimated from
a dynamic process starting from 
an ordered state \cite {luo97a}.
The results are in good agreement with and
improve the measurements in equilibrium.
For the XY model, taking the exponents $\theta$ 
and $\eta$ obtained in Refs.~\cite {oka97,luo97a}
as input, one can compute
separately the dynamic exponent $z$ from $\lambda$ and $y$ 
for different temperatures.
The results are given in Table~\ref {t1}.
The values of $z$ obtained from the auto-correlation and 
the second moment agree well within statistical errors.
This fact strongly supports the scaling forms in
Eqs.~(\ref {e40}) and (\ref {e60}) and the scaling relations
between the exponents in Eq.~(\ref {e80}) and (\ref {e180}).
However, the dynamic exponent $z$ is clearly different from
that measured from a dynamic process quenched 
from an ordered initial state, which is denoted by $z_1$ in 
Table~\ref {t1}.
As the temperature decreases,
$z_1$ remains a constant near $2$, but $z$
is apparently bigger than $2$ and increases
gradually. The difference is $15$ to $20$ percent
and we regard it as prominent.
This indicates that the dynamic scaling behaviour
of the XY systems with a Kosterlitz-Thouless phase transition
is more complicated than the simple scenario of the dynamic 
scaling
for critical systems with second order phase transitions.
The dynamic XY model has a non-trivial crossover behaviour
for the time $t$ macroscopically not small but also not asymptotically large.
As mentioned before, the microscopic time scale
$t_{mic}$ grows as the temperature decreases.
When the temperature approaches zero, 
it might happen that the system evolves
already into the crossover regime at the time $t \sim t_{mic}$,
and thus the simple scaling forms (\ref {e40}) and (\ref {e60}) are not 
observed.

For the FFXY model, the situation is even more complicated.
Taking $\eta$ as input, the resulting dynamic exponent $z$
from the second moment interestingly shows a non-monotonous
dependence on the temperature. This was also
qualitatively observed by other authors in calculating the 
domain growth \cite {lee95}. However, if we take
the exponent $\theta$ as input, the dynamic exponent
estimated from the auto-correlation with the scaling relation
(\ref {e80}) does not coincide with that from 
the second moment. The difference can not be explained by
the statistical errors. Since the exponent $y$ from the second 
moment
is not directly related to
the initial conditions,
we believe that the scaling relation (\ref {e180})
holds in any case, i.e. the exponent $z$ computed from 
$y$ should be correct.
However, for the FFXY model the scaling relation (\ref {e80})
is violated. The violation can be described by
$\delta=\lambda-d/z+\theta$. Values for this quantity are given in
Table~\ref {t2}. The origin of this violation 
may probably be traced back to the chiral degree of freedom.
Since the chiral transition temperature $T_c$ is above $T_{KT}$,
 ordering dynamics of the chiral magnetization may affect
 the behaviour of the auto-correlation. 
 However, a complete understanding of this problem remains 
open.

In conclusions, we have numerically investigated
the short-time dynamic 
 behaviour of the two-dimensional XY and the fully frustrated XY 
model
 at the temperature $T_{KT}$ and below.
 Our results show that there exists
 indeed dynamic scaling in the short-time regime.
 However, the dynamic exponent $z$ may depend on the initial
 conditions. It indicates that a non-trivial crossover
 behaviour occurs when the dynamic evolution crosses over from 
macroscopic
 short-time regime to the long-time regime.
 This scenario is very different from
 that of critical systems with second order phase transitions.

\vspace{0.2cm}

{\bf Acknowledgements}:
Work supported in part by the Deutsche 
Forschungsgemeinschaft;
Schu 95/9-1 and SFB~418.

\begin{table}[h]\centering
\begin{tabular}{cllll}   
 $T$         & 0.90     & 0.86      &  0.80      & 0.70  \\
\hline
 $\lambda$   & 0.625(4) & 0.600(3)  &  0.569(2)  & 0.552(4) \\
$y$          & 0.766(16)& 0.775(15) &  0.780(19) & 0.773(24) \\
\hline
$\theta$     & 0.250(1) & 0.264(5)  &  0.290(1)  & 0.287(3) \\
$\eta$       & 0.244(5) & 0.212(4)  &  0.178(2)  & 0.143(3)\\
\hline
$z=(2-\eta)/y$       
             & 2.292(48)& 2.307(45) &  2.335(57) & 2.402(75)\\
$z=d/(\theta+\lambda)$ 
             & 2.286(11)& 2.315(16) &  2.328(06) & 2.384(14) \\
\hline
 $z_1$       & 1.96(4)  & 1.98(4)   &  1.94(2)   & 1.98(4) \\
\hline
\end{tabular}
\caption{The exponents $\lambda$ and $y$ measured for the XY 
model.
Values of the exponent $\theta$ are taken from
Ref.~\protect\cite {oka97}, while
 $\eta$ and $z_1$  are from 
Ref.~\protect\cite {luo97a}. $z_1$ is the dynamic exponent $z$
measured from a quench from an ordered initial state
\protect\cite {luo97a}.
}
\label{t1}
\end{table}

\begin{table}[h]\centering
\begin{tabular}{cllll}   
 $T$         &\ \ \ 0.44     &\ \  0.40      &\ \   0.35      &\ \  0.30  \\
\hline
 $\lambda$   &\ \ \  0.808(3) &\ \   0.622(2)  &\ \    0.577(2)  &\ \   0.561(2) \\
$y$          &\ \ \  0.747(14)&\ \   0.835(20) &\ \    0.809(15) &\ \   0.801(13) \\
\hline
$\theta$     &\ \ \  0.079(4) &\ \   0.181(5)  &\ \    0.245(3)  &\ \   0.263(2) \\
$\eta$       &\ \ \  0.243(4) &\ \   0.140(2)  &\ \    0.107(1)  &\ \   0.086(2)\\
\hline
$z=(2-\eta)/y$       
             &\ \  \ 2.352(44)&\  \ 2.228(53)  &\ \ 2.340(43)&\ \ 2.390(39)\\
$\delta=\lambda-d/z+\theta$ 
             &\ \ \ 0.037(17)&  $-$0.095(22) &  $-$0.033(16)   &  $-$0.013(14) 
\\
\hline
 $z_1$       &\ \ \ 1.93(4)  &\ \ 1.95(3)   &\ \  1.99(7)   &\ \ 1.97(3) \\
\hline
\end{tabular}
\caption{The exponents $\lambda$ and $y$ measured for the 
FFXY model.
Values of the exponents $\theta$ are taken from 
Ref.~\protect\cite {luo98a}, while  
 $\eta$ and $z_1$  are from 
Ref.~\protect\cite {luo98a}. $z_1$ is the dynamic exponent $z$
measured from a quench from an ordered initial state
\protect\cite {luo97a}.}
\label{t2}
\end{table}

\begin{thebibliography}{10}

\bibitem{jan89}
{H. K. Janssen, B. Schaub and B. Schmittmann}, Z. Phys. {\bf {B 73}} (1989)
  539.

\bibitem{hus89}
D.~A. Huse, Phys. Rev. {\bf {B 40}} (1989) 304.

\bibitem{sta92}
D. Stauffer, Physica {\bf {A 186}} (1992) 197.

\bibitem{li94}
{Z.B. Li, U. Ritschel and B. Zheng}, J. Phys. A: Math. Gen. {\bf {27}} (1994)
  L837.

\bibitem{sch95}
L. {Sch\"ulke} and B. Zheng, Phys. Lett. {\bf {A 204}} (1995) 295.

\bibitem{zhe98}
B. Zheng, Int. J. Mod. Phys. {\bf B12} (1998) 1419, review article.

\bibitem{luo98}
{H.J. Luo, L. Sch\"ulke and B. Zheng}, Phys. Rev. Lett. {\bf {81}} (1998) 180.

\bibitem{li96}
{Z.B. Li, L. {Sch\"ulke} and B. Zheng}, Phys. Rev. {\bf E 53} (1996) 2940.

\bibitem{sch96}
L. Sch{\"u}lke and B. Zheng, Phys. Lett. {\bf {A 215}} (1996) 81.

\bibitem{gra95}
P. Grassberger, Physica {\bf {A 214}} (1995) 547.

\bibitem{blu92}
{R. E. Blundell, K. Humayun and A. J. Bray}, J. Phys. A: Math. Gen. {\bf {25}}
  (1992) L733.

\bibitem{jan92}
{H. K. Janssen},  in {\em {From Phase Transition to Chaos}}, edited by {G.
  Gy\"orgyi, I. Kondor, L. Sasv\'ari and T. T\'el, Topics in Modern Statistical
  Physics} (World Scientific, Singapore, 1992).

\bibitem{luo97a}
{H.J. Luo and B. Zheng}, Mod. Phys. Lett. {\bf B11} (1997) 615.

\bibitem{rut95}
A.~D. Rutenberg and A.~J. Bray, Phys. Rev. {\bf {E 51}} (1995) R1641.

\bibitem{blu94}
{R. E. Blundell and A. J. Bray}, Phys. Phys. {\bf {E49}} (1994) 4925.

\bibitem{oka97}
{K. Okano, L. {Sch\"ulke}, K. Yamagishi and B. Zheng}, J. Phys. A: Math. Gen.
  {\bf 30} (1997) 4527.

\bibitem{luo98a}
{H.J. Luo, L. Sch\"ulke and B. Zheng}, Phys. Rev. {\bf {E57}} (1998) 1327.

\bibitem{sch89}
A. Scheinine, Phys. Rev. {\bf {B 39}} (1989) 9368.

\bibitem{lee95}
{S. J. Lee, J. R. Lee and B. Kim}, Phys. Rev. {\bf {E 51}} (1995) R4.

\end{thebibliography}
\end{document}